\documentclass[a4paper,noarxiv, twocolumn,11pt]{quantumarticle}
\pdfoutput=1
\usepackage[utf8]{inputenc}
\usepackage[english]{babel}
\usepackage[T1]{fontenc}
\usepackage{amsmath}
\usepackage{hyperref}
\usepackage{graphicx}
\usepackage{subfigure}
\usepackage{mathtools}
\usepackage{empheq}
\usepackage{multirow}
\usepackage{color}
\usepackage[table,x11names]{xcolor}
\usepackage{tikz}
\usepackage{lipsum}
\usepackage[square,numbers,sort&compress]{natbib}

\begin{document}

\title{Matter-wave interferometry using a levitated magnetic nanoparticle}

\author{A T M Anishur Rahman}
\affiliation{Department of Physics \& Astronomy, University College London, Gower Street, WC1E 6BT, UK}
\email{a.rahman@ucl.ac.uk}

\maketitle
\begin{abstract}
  The superposition principle is one of the bizarre predictions of quantum mechanics. Nevertheless, it has been experimentally verified using electrons, photons, atoms, and molecules. In this article, using a $20~$nm levitated ferromagnetic FePt nanoparticle, an exotic all optical spin polarization technique and the matter-wave interferometry, we show that a mesoscopic spatial Schrodinger cat can be created. Additionally, we argue that the maximum spatial separation between the delocalized wavepackets can be $25~\mu m$ and is significantly larger than the object itself.  
\end{abstract}

\section{Introduction} 
Quantum mechanics prescribes \cite{Arndt2014} that, irrespective of their size or mass, an object can be in multiple states at once and this is certainly true for microscopic systems \cite{Monroe1996,Arndt1999,Eibenberger2013,Arndt2014,FeinYaakov2019NatPhys}. Specifically, in the past, the quantum superposition principle has been demonstrated using neutrons \cite{ZawiskyM2002}, electrons \cite{Arndt2014}, ions \cite{Monroe1996} and molecules \cite{Arndt1999,Eibenberger2013}. In 1996, Monroe et al. created a spatial Schrodinger cat state in which they put a beryllium ion in two different spatial locations separated by 80 nm at the same time \cite{Monroe1996}. The size of the beryllium ion was approximately $0.1~$ nm in size ($\approx 1.5\times 10^{-26}~$kg). In another experiment \cite{FeinYaakov2019NatPhys}, oligoporphyrins molecules with high kinetic energy were sent through different grating structures and the resulting matter-wave interfered after a free flight due to the wave nature of matter. In this case, the wave packet of oligoporphyrins was delocalized by $266~$nm. The current record for the largest spatial superposition is $0.5~$m which was realized using a Bose-Einstein condensate of Rubidium atoms in an atomic fountain \cite{Kovachy2015}, while the heaviest object so far put into a superposition state is about $4.446425\times 10^{-23}~$ kg \cite{FeinYaakov2019NatPhys}.

Increasing the macroscopicity of a spatial superposition state is an ongoing global effort and many proposals have been put forward using clamped and levitated optomechanical systems \cite{BosePRA1999,Marshall2003,Isart2011,Zhang2013,BatemanNatComm2014,WanPRL2016,IsartNJP2016,Rahman_2019}. In these schemes the spatial separations between the two arms of the superposed states are less than the size of the respective objects ($\approx 100~$nm ). Nevertheless, a successful experimental demonstration of such a state using a mesoscopic system can resolve the apparent conflict between quantum mechanics and general relativity \cite{Arndt2014}. A demonstration of the quantum superposition principle using a mesoscopic object can also shed light on the gravity's role on the quantum state reduction \cite{Penrose1996OGri} and test different wavefunction collapse models \cite{Bassi2013,Carlesso2019}.

In this article, we theoretically show that the spatial separation of a quantum superposed state can be increased significantly by using existing technologies. To achieve this, we exploit recent progress, the optical spin polarization of ferrimagnetic or ferromagnetic materials \cite{HohlfeldAPL2009,StanciuPRL2007}. In particular, we use all optical helicity dependent switching for flipping spin states of ferrimagnetic or ferromagnetic materials at cryogenic temperature. Exploiting this helicity dependent spin polarization, a pragmatic magnetic field gradient, and a levitated nanosphere, we show that, in principle, a separation of at least $25~\mu m$ between the delocalized superposed states can be achieved.

\begin{figure}
		\includegraphics[width=8.50cm]{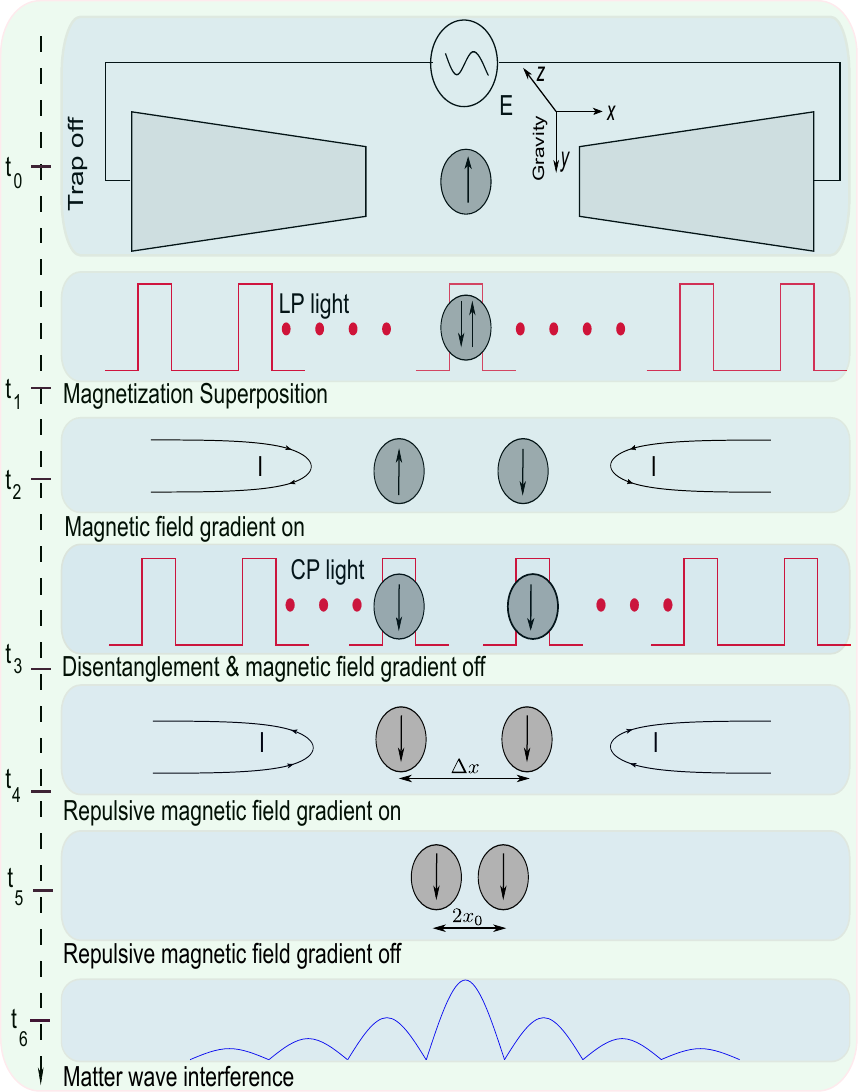}
		\caption{A FePt nanoparticle is levitated using an ion trap at a cryogenic temperature of $300~$mK. Subsequently, using all optical helicity dependent switching and a magnetic field gradient, a spatial superposition is created. In creating and detecting the spatial superposition following main steps are followed (see main text for more information) - (i) ion trap off, (ii) creation of magnetization superposition using linearly polarized (LP) light, (iii) magnetic field gradient on, (iv) magnetic field off, (v) disentanglement using circularly polarized (CP) light, (vi) repulsive magnetic field to bring back the matter waves towards the centre of the trap where they interfere to reveal spatial superposition created earlier, v) Repulsive magnetic field is off, and vi) Matter wave interference pattern after the free evolution of the wavepackets.}
	\label{fig1}
\end{figure}


\section{Spatial superposition} A schematic of the proposed experiment and the sequence of events required to create the spatial superposition are shown in Fig. \ref{fig1}. In the proposed scheme, a single domain ferromagnetic or ferrimagnetic particle is levitated using an ion trap at a cryogenic temperature of $300~$mK \cite{AldaAPL2016,Huillery2019,VinantePRA2019,Bykov2019,Bullier_2020}. Other platforms of levitation include magnetogravitational, optical and magnetic trap \cite{Gieseler2019,VinanteA2020Umdw,Slezak_2018,RahmanNatPhot2017,SebersonT2019Oloa,RahmanSLD2020}. Spins in a single domain ferrimagnetic or ferromagnetic nanoparticle are strongly interacting, highly correlated and perfectly ordered in the ground state due to the strong exchange coupling among the electrons \cite{AwschalomScience1992,TabuchiPRL2014}. Consequently, the spins of a small single domain ferrimagnet or ferromagnet can be considered as a single superspin and can be described using magnetization $M=g_l\mu_BS$ \cite{AwschalomScience1992,ChudnovskyPRL2000,TabuchiPRL2014,RusconiPRL2017}, where $g_l$ the Lande factor, $\mu_B$ is the Bohr magneton and $S$ is the total uncompensated spin. After levitation, the centre of mass (CM) temperature $T_{cm}$ of the nanoparticle is reduced down to a suitable temperature for a better visibility in the matter-wave interferometry \cite{Bykov2019,AldaAPL2016}. Here, the motional ground state of the particle is ideal but not necessary (see below).

%
%
%

Once the required CM temperature is achieved, the ion trap is switched off in such a way that the particle remains unperturbed. Subsequently, using all-optical helicity dependent switching (AOHDS), a superposition of magnetization (spin) is created. Specifically, it has been demonstrated that the magnetization of certain ferrimagnetic (GdFeCo) and ferromagnetic (FePt,CoPt) thin films, composed of many domains, change from a down $|\downarrow_{M}\rangle$ (up $|\uparrow_{M}\rangle$) polarized state to an up $|\uparrow_{M}\rangle$ (down $|\downarrow_{M}\rangle$) polarized state when exposed to a femtosecond right $|R\rangle$ (left $|L\rangle$) circularly polarized light. In contrast, when the same film is exposed to linearly polarized light ($(|R \rangle+|L \rangle)/\sqrt{2}$) pulses, the magnetization of each domain becomes either up or down polarized at random \cite{StanciuPRL2007,LambertScience2014,ManginNatMat2014,JohnRSciRep2017}. In another word, linearly polarized light creates a superposition of all up and all down spin states but the superposition collapses to one of the eigenstates - possibly due to the simultaneous measurements that these experiments undertake to determine the effect of the laser pulse on the magnetic film.

Although the exact mechanism behind AOHDS is not clearly understood yet \cite{ElHadriPhysD2017,ZhangGRev2016}, the combined effect of the laser induced heating and the effective magnetic field associated with the laser pulse, the inverse Faraday effect, and an enhanced spin-orbit coupling are among the mechanisms believed to be responsible for this phenomenon \cite{StanciuPRL2007,ZhangGRev2016,ElHadriPhysD2017}. In a later experiment \cite{HohlfeldAPL2009}, it has been demonstrated that AOHDS works better at a cryogenic temperature. Indeed, it has been shown that a lower laser fluence than the original room temperature experiment \cite{StanciuPRL2007} is sufficient to change the spin state. Additionally, in these experiments \cite{StanciuPRL2007,LambertScience2014,ManginNatMat2014,JohnRSciRep2017}, it has been found that the number of photons required is significantly less than the number of spin flipped \cite{StanciuPRL2007,HohlfeldAPL2009}. For example, in the first experiment \cite{StanciuPRL2007}, the number of photons used to flip $\approx 1\times 10^{13}$ spins in $20~\mu m$ film was $\approx 1\times 10^{11}$. This also negates the possibility of a spin flip due to a single photon absorption.


To create a spatial superposition, once the ion trap has been switched off and a spin superposition has been created, an inhomogeneous dc magnetic field \cite{Harrison2015} is activated at time $t_2$ and the particle evolves under the influence of gravitational and magnetic fields for a suitable time $t=t_3-t_2$ (see Fig. \ref{fig1}). The spatial separation \cite{WanPRL2016,Rahman_2019} between the two arms of a superposed states after time $t$ is

\begin{eqnarray}
\Delta x =\frac{M t^2}{m}\frac{dB}{dx}, 
\end{eqnarray}

where $M=g_l \mu_B S_x$ is the magnetization of the particle, $S_x$ is the spin projection along the quantization axis ($x-$axis), $g_l$ is the Lande factor and $\mu_B$ is the Bohr magneton. $dB/dx$ is the magnetic field gradient and $m$ is the mass of the levitated particle. 


\textbf{Detection:} After the desired separation $\Delta x$ between the superposed states $|\phi\rangle=(|\uparrow_{M}\rangle|\phi_L(-x_l)\rangle +|\downarrow_{M}\rangle|\phi_R(x_r)\rangle)/\sqrt{2}$ is achieved, where $|\phi_L(-x_l)\rangle$ and $|\phi_R(x_r)\rangle$ are the spatial states at position $-x_l$ and $x_r$, respectively, a disentangling circularly polarized light pulse is activated and the initial magnetic field gradient is switched off. A right (left) circularly polarized light pulse reduces a superposed state into an up (down) spin state. The state of the system at this stage is $|\phi_S\rangle=(|\phi_L(-x_l)\rangle +|\phi_R(x_r)\rangle)/\sqrt{2}$. Note that once the spin superposition has been reduced to one of the eigenstates via the disentangling operation, the experiment is no longer limited by the finite spin coherence time. Instead, the experiment is restricted by the centre-of-mass coherence time which can be very long in ultra high vacuum \cite{Chang2010}. Incidently, a cryogenic experimental condition, as aimed in this proposal, inherently provides this level of vacuum \cite{VinantePRA2019} and hence a long motional coherence time is naturally guaranteed. Nevertheless, to reduce the free evolution time, at this stage of the experiment, a new inhomogeneous dc magnetic field is activated which redirects the separated wave packets towards the centre. Once the wavepackets approach each other from opposite directions, the polarity of the magnetic field is changed which decelerates the wavepackets. Note that the pulsed operation of a dc electromagnet with $500~$ns pulse has been recently demonstrated \cite{Harrison2015}. After waiting for an adequate time when the terminal velocities of the wavepackets approach zero, the magnetic field is switched off. Subsequently, at the centre, the two wavepackets interfere to produce an interference pattern. Specifically, let the two arms of the superposed states evolve in free space for an additional time $t_f=t_6-t_5$. After this free flight, the two arms of the matter-wave interfere \cite{Arndt1999,Zhang2013,BatemanNatComm2014} on a substrate just like in Young's double slit experiment. Overall, the sequence of events described above is repeated many times to reveal the fringe pattern. Eventually, this interference pattern is used as the signature of the spatial superposition created. The period of the interference pattern \cite{Zhang2013} is given by $P_0=2\pi\hbar t_f/(m x_0)$, where $m$ is the mass of the particle and $\pm x_0$ are the positions of the wavepackets when their velocities were zero.

\section{Results and discussion}
The visibility of the matter-wave fringe pattern depends on, among other factors, the position uncertainty of the levitated nanoparticle before the ion trap is switched off. An uncertainty in the position translates into an uncertainty of $x_0$. Here, the motional ground state of the particle can be useful. However, it is not necessary as long as $\sqrt{<x^2>}<<P_0$, where $x$ is the particle's instantaneous position. This seems readily achievable given the sub-kelvin ambient temperature and the ultra high vacuum that the cryogenic environment provides \cite{VinantePRA2019}. Additionally, due to the small particle size ($10~$nm radius) considered in this article, the secular frequency of the oscillator \cite{Bullier_2020} $\omega=\frac{q_e e V_0}{\sqrt{2}m\Omega_d r_0^2}$ can be significantly higher than usual \cite{Bullier_2020} which naturally leads to a lower uncertainty in the position. Here, $q_e$ is the number of elementary charge and $e$ is the magnitude of the electronic charge. $V_0~$ and $\Omega_d$ are the amplitude and the frequency of the applied ac voltage. $r_0$ is the distance of the trap centre from the electrodes and $m$ is the mass of the particle. For example, for $q_e=5$, $m=2\times 10^{-20}~$kg, $r_0=1~$mm, $\Omega_d=10~$kHz and $V_0=300~$V, one gets $\omega/2\pi\approx 150~$kHz. For $T_{cm}=1~$mK, we find $\sqrt{<x^2>}\approx 1~$nm. This is significantly lower than $P_0\approx 60~$nm (see below). Hence, the visibility of the fringe pattern is robust against the uncertainty in the position. Radiation pressure, originated from the laser pulse used in the creation of spin superposition, can be significant. For an estimate, if we consider $1\times 10^4$ photons (actual number can be different) approximately required for about $1\times 10^6$ spins in particle then the resultant velocity of the particle due to photon recoil is $\approx 5\times10^{-4}~m~s^{-1}$ or equivalently $T_{cm}=600~$mK. This is not negligible. However, this is a deterministic effect and hence cannot reduce the visibility of the interference fringe but can shift the fringe pattern away from the centre. This can be compensated by using magnetic fields of different amplitude from each side in the last step (see Fig. \ref{fig1}). Polydispersity of nanoparticles, responsible for different de Broglie wavelengths, is another factor which can reduce the visibility of the fringe pattern. Fortunately, nanoparticles used in this proposal (FePt) have been synthesized in monodisperse configuration \cite{SunScience2000} and consequently, the current scheme is robust.


\textbf{Spin coherence time:} One of the main requirements of this experiment is the coherence of many spins contained in a ferromagnetic nanoparticle. Spin coherence time is characterized by using two time constants i.e. the spin-lattice relaxation time $T_1$ and the transverse relaxation time $T_2$, and in general $T_2 \le 2T_1$\cite{Zutic2004}. While $T_1$ is determined by the spin-lattice interaction, $T_2$ is primarily fixed by the inhomogeneities and fluctuation in magnetic and crystal fields. For homogeneous isotropic material $T_1=T_2$ \cite{Zutic2004,BlochPR1946}. In ferromagnetic materials, the usage of $T_1$ and $T_2$ is not prevalent rather Gilbert damping $\alpha$, which characterizes how a spin system loses energy and angular momentum, is widely used \cite{Beausjour2009,SchoenMaw2016,CapuaAmir2017}. Nevertheless, $\alpha$, $T_1$ and $T_2$ are intricately related with each other \cite{CapuaPRB2015,CapuaAmir2017}. In the absence of inhomogeneities, valid for small nanocrystals, coherence time is given by $T_2=1/(\alpha \gamma B)$ \cite{CapuaPRB2015}, where $B$ is the magnetic field and $\gamma$ is the gyromagnetic ratio. In bulk FePt sample, $\alpha$ $ <1\times 10^{-2}$ has been measured \cite{Iihama2013,Fuller2009}. Since a single domain $20~$nm FePt nanocrystal is considered in this article, a smaller $\alpha$ and hence a longer coherence time can be expected. Levitation, a physical contactless low noise environment, may boost the coherence time further. Additionally, performing the experiment at a subkelvin temperature will increase the coherence time significantly. A cryogenic temperature is also beneficial for suppressing magnons \cite{Coey2009} which is another mechanism that shortens the coherence time. It is also important to mention that due to the finite size of the magnetic nanoparticles considered in this article, propagating magnons will be heavily damped and only the high energy oscillations can be excited \cite{TabuchiPRL2014}.

\textbf{Example:} Although all optical helicity dependent switching has been demonstrated using many different materials, in this proposal we use crystalline ferromagnetic face centred tetragonal (L1$_0$) FePt. It has one of the highest magnetocrystalline anisotropy and is readily available in monodisperse single domain nanocrystalline form \cite{SunScience2000}. All optical helicity dependent switching using a granular film of FePt nanoparticles has also been recently demonstrated \cite{JohnRSciRep2017} using this material. Ferromagnetism, a manifestation of quantum exchange coupling, of FePt guarantees that all the spins contained within a levitated particle are exchange-coupled and hence behave as a single coherent macrospin. FePt is also preferable due to most of the naturally abundant isotopes of iron ($\approx 98\%$) and platinum ($\approx 70\%$) are nuclear spin free which guarantees enhanced spin coherence time \cite{BlochPR1946,Zutic2004}.

\begin{figure}
	\includegraphics[width=8.5cm]{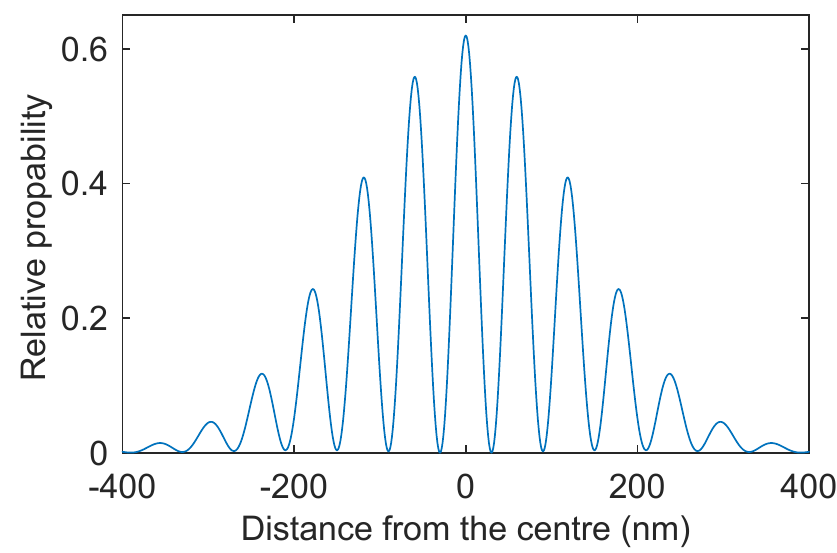}
	\caption{Matter wave interference \cite{Zhang2013} after a free-flight of $t_f=100$ ms. Parameters used are $r=10$ nm, $T_{cm}=1~$mK, and the separation between the two arms is $x_0=20$ nm when the velocities of the wavepackets approaching each other from the opposite directions are zero.}
	\label{fig2}
\end{figure}

To provide an example, let us consider a FePt particle of radius $r=10$ nm levitated and prepared in spatial superposition as described before. Assuming a saturation magnetization $M_s=1\times 10^6~$$A~m^{-1}$ \cite{Iihama2013} with $M=VM_s$ and $V$ is the volume of the particle, $\frac{dB}{dx}=1\times 10^{3}~$$T~m^{-1}$ \cite{Harrison2015}, $\rho=4000~$kg~m$^{-3}$ and $t=T_2=10~$$\mu$s, one gets $\Delta x\approx 25~\mu m$. This is a macroscopic distance. Note that the spin coherence time used in this calculation is an estimate only. In fact, a measured value of the spin coherence time of FePt is not available in literature. In retrospect, measuring the spin coherence time of a ferromagnetic nanoparticle will be of great interest for many applications. Figure \ref{fig2} shows the relevant matter-wave interference pattern \cite{Zhang2013} after the free evolution of the wavepackets for $t_f=100~$ms. The period of this fringe pattern is $P_0\approx 60~$nm and can be changed by controlling $x_0$. $P_0$ can be made larger by having a longer $t_f$ but this will increase the physical size ($gt_f^2/2$) of the experiment, where $g$ is the gravitational acceleration. The interference pattern can be post processed using an electron microscope to physically reveal the spatial superposition created. This, however, cannot show the maximum spatial separation that the current scheme can produce. For that, in some run of the experiment, one can perform measurements just before the disentangling pulse is applied (see Fig. \ref{fig1}). Off-course this will collapse the wavefunction. But this is essential to reveal the advantage of the current scheme. 

\textbf{Macroscopicity:} In the scale of macroscopicity $\mu=log_{10}\Bigl[\Bigl | \frac{1}{\ln{f}}(\frac{m}{m_e})^2\frac{t}{1~s}\Bigr |\Bigr]$ \cite{NimmrichterPRL2013}, the combination of the particle mass and the spin coherence time used in the previous example amount to $\approx 16$, where $f$ is the fringe visibility, $m$ is the mass of the object, $t$ is the spin coherence time and $m_e$ is the electronic mass. In the calculation, we have used $f=1$. At present the state of the art experiment has $\mu=14$ \cite{FeinYaakov2019NatPhys}.

\section{Conclusions}
We have theoretically shown that using a levitated single domain magnetic nanoparticle and all optical helicity dependent switching, a spatial superposition state can be created. In this scheme, the spatial separation between the two arms of a superposition state is $\approx 25~\mu m$. This is significantly larger than the actual object put into the superposition. However, there are significant experimental challenges to overcome. One such challenge is the creation of the spin superposition using the all optical helicity dependent switching. While many experiments have demonstrated this phenomenon in the classical context, it remains to be seen in the quantum realm. Likewise, the spin coherence time in ferromagnetic systems containing many spins has not been measured yet. Indeed, measuring the spin coherence time in these systems will itself be very interesting. This can potentially open up many new applications in spintronics and other spin based systems.


\bibliographystyle{unsrtnat}
\bibliographystyle{abbrvnat}

\end{document}